\documentclass[aps,a4paper,onecolumn,11pt,unpublished]{quantumarticle}
\pdfoutput=1

\usepackage{amsmath,amssymb}
\usepackage{booktabs}
\usepackage{array}
\usepackage{graphicx}
\usepackage{placeins}
\usepackage{xcolor}
\usepackage{xurl}
\usepackage{hyperref}
\newcommand{\dF}{\Delta\mathrm{F1}}
\newcommand{\dA}{\Delta\mathrm{ACC}}
\newcommand{\pp}{\,\mathrm{pp}}

\begin{document}

\title{HyQ-LC: Improving Light-Curve Classification with Hybrid Quantum--Classical Learning}

\author[orcid=0009-0000-4773-6574]{Xinghang Yu}
\email{yuxinghang@tju.edu.cn}
\affiliation{School of Computer Science and Technology, Tianjin University, 135 Yaguan Road, Haihe Education Park, Tianjin 300350, People's Republic of China}
\author[orcid=0000-0002-3783-2228]{Bin Yang}
\affiliation{School of Computer Science and Technology, Tianjin University, 135 Yaguan Road, Haihe Education Park, Tianjin 300350, People's Republic of China}
\email{yangbincic@tju.edu.cn}

\begin{abstract}
Light curves reveal stellar variability and support studies of stellar evolution and cosmic distances. Reliable classification is therefore important for large time-domain surveys. We present HyQ-LC for ten-class stellar light-curve classification using 359 handcrafted features and 633{,}034 training objects. The main model combines a window multilayer perceptron (MLP) backbone with a nine-qubit readout using amplitude encoding and angle re-uploading. Both this hybrid and its pure-MLP control have 44{,}238 parameters. On a shared 20{,}000-object test set, the hybrid has macro-averaged F1 (Macro-F1) point gains of 0.8276 and 1.2393 percentage points over the pure MLP on two fine-tuning seeds. A separate fixed-fusion configuration achieves slightly higher Macro-F1 but lower accuracy than approximately budget-matched LightGBM. Its positive class-level differences are concentrated in CEP, RRC, and ROT, while the full-budget LightGBM reference scores higher on both metrics. All circuits are simulated classically. The comparisons retain existing validation-selected checkpoints and use the same test objects, with features, predictions, and recomputation code publicly available. The results identify point-estimate improvements over the tested controls while leaving the specific contribution of the quantum branch unresolved.
\end{abstract}

\maketitle

\section{Introduction}

Time-domain surveys generate large collections of light curves that require scalable classification~\cite{moriwaki2023}. Feature-based systems such as ALeRCE~\cite{alerce2021} and incremental frameworks such as TDLight~\cite{tdlight2026} address this task through complementary classification and data-management strategies. Learned representations provide another route, including ASTROMER embeddings~\cite{astromer2023}, time-series foundation models benchmarked by StarEmbed~\cite{starembed2026}, and structured sequence modelling with large language models~\cite{shao2026}. These studies provide context; their different inputs, labels, and evaluation protocols prevent direct score comparisons with the present task.

Hybrid quantum--classical learning combines classical processing with trainable quantum representations~\cite{mari2020}. Peters et al. demonstrated quantum-kernel classification of supernova spectral features on a noisy processor~\cite{peters2021}. Architecture comparisons on tabular data also motivate studying how quantum branches interact with classical networks~\cite{chenkuo2026}. However, expressive circuits alone do not guarantee predictive improvements. Bowles et al. showed that classical baselines and entanglement-removal controls can substantially change the interpretation of quantum-learning benchmarks~\cite{bowles2024}.

The distinction between a useful hybrid classifier and a useful quantum branch is central to this evaluation. A hybrid can improve on a dense network because of its feature topology, pre-training, or residual head, even when a classical replacement supplies similar information. Parameter matching controls model size but does not remove these alternatives. Handcrafted light-curve features also define a tabular learning problem, making a strong tree-based reference relevant alongside neural controls.

Here we ask whether a hybrid representation improves stellar light-curve classification under controlled parameter budgets. We distinguish comparisons with a parameter-matched pure MLP from replacements of the quantum branch within a shared network. A separate fixed-fusion configuration supports comparisons with classical fusion and an approximately budget-matched LightGBM. Every reported test comparison uses the same 20{,}000 objects.

\section{Models and Evaluation}

\subsection{Data and task formulation}

We use handcrafted features derived from the LEAVES light-curve dataset~\cite{leaves2024,leavesdata}. The task contains nine variable-star categories and one non-variable category. The labels are CEP, DSCT, EA, EW, M, Non-var, ROT, RRAB, RRC, and SR; Non-var denotes non-variable objects. Each object has 359 precomputed handcrafted features. The training pool contains 633{,}034 objects, and the reported test cohort contains 20{,}000 objects with distinct feature-bank row IDs. These released IDs are not original survey identifiers. EW and SR together account for 61.89\% of the test cohort; CEP has only 45 objects. Class counts and classwise scores are listed in Table~\ref{tab:classwise} and provided with the released data~\cite{artifacts}.

Preprocessing uses frozen medians and scales estimated from an original 50{,}000-object training subset, median replacement of non-finite values, and clipping to $[-10,10]$. The main models use 105{,}297 validation objects; the fixed-fusion and leaf-budget LightGBM experiments use 10{,}000. We retain their existing validation-selected checkpoints and hyperparameters. Test predictions are extracted by exact ID and label matching from saved prediction files. The cohort was previously examined during exploratory development, so it is not an untouched confirmation sample. No separate confirmation cohort is evaluated here.

We formulate classification as a map from a preprocessed feature vector $\mathbf{x}_i\in\mathbb{R}^{359}$ to a label $y_i\in\{1,\ldots,C\}$, with $C=10$. The networks receive engineered features rather than raw sequences of observation times and magnitudes. Consequently, the windows below operate over entries of the feature vector; they do not denote consecutive time intervals in a light curve. This common input representation allows direct comparisons with dense MLPs and gradient-boosted trees.

\subsection{Composition of the 359-feature representation}

The input is a fixed statistical feature representation of each light curve, organised into seven disjoint groups (Table~\ref{tab:featuregroups}). Their dimensions sum to $81+17+32+49+63+64+53=359$. We retain the complete input and its original column order for every model. The ordered feature names and group assignments are released as \texttt{data/feature\_manifest.csv}~\cite{artifacts}.

The base group contains 81 descriptors: mean and standard deviation; four families of variability statistics at 16 time scales from 0.02 to 655.36 days; recent mean and variance at four time scales; and seven additional variability descriptors. The additional 17 global descriptors capture distribution moments, trends, sampling, and measurement errors. The 32 histogram-derived descriptors summarise quantiles, robust widths, and tail shape. These histogram summaries approximate distribution information rather than retaining an exactly sorted light curve.

Local temporal structure is represented in two complementary ways. At observation-index lags $1,2,4,8,16,32,64$, seven pair descriptors give 49 features, and six ordinal patterns plus three curvature descriptors give 63 three-point features. Another 64 features summarise differences in 16 physical-time bins. Index lags measure separation in observation order; physical-time bins describe elapsed time, which varies under irregular sampling.

The final 53 features describe a cached candidate-frequency pool. They contain six descriptors for each of four leading peaks, three for each of eight frequency bands, and five pool-level summaries, giving $4\times6+8\times3+5=53$. These include candidate periods, powers, fundamental and second-harmonic amplitudes, harmonic ratios, and power-distribution summaries. They encode periodicity evidence from the candidate pool rather than asserting recovery of the true stellar period. The present study evaluates classifiers on the resulting feature vectors; it does not evaluate the online update cost of their extractor.

\begin{table}[htbp]
\centering
\small
\setlength{\tabcolsep}{4pt}
\caption{Composition of the shared 359-dimensional input. Groups partition the feature vector; all columns are retained. The released manifest records the exact zero-based index, feature name, and group.}
\label{tab:featuregroups}
\begin{tabular}{@{}>{\raggedright\arraybackslash}p{0.27\linewidth}r>{\raggedright\arraybackslash}p{0.62\linewidth}@{}}
\toprule
Feature group & Dim. & Representative information \\
\midrule
Base and time-scale statistics & 81 & Mean, standard deviation, multiscale variability, recent statistics, and adjacent changes. \\
Global, sampling, and error statistics & 17 & Skewness, kurtosis, trends, observation count/span, and measurement-error descriptors. \\
Histogram distribution summaries & 32 & Approximate quantiles, robust widths, median, tail shape, and asymmetry. \\
Observation-lag pair statistics & 49 & Seven descriptors at each of seven observation-index lags. \\
Three-point temporal patterns & 63 & Six ordinal patterns and three curvature descriptors at each of seven lags. \\
Physical-time-bin differences & 64 & Four difference descriptors in each of 16 physical-time bins. \\
Cached periodicity descriptors & 53 & Candidate periods, powers, harmonic amplitudes, and frequency-pool summaries. \\
\midrule
Total & 359 & The same input representation is used by all compared models. \\
\bottomrule
\end{tabular}
\end{table}

\subsection{Classical backbone and model overview}

\begin{figure}[t]
\centering
\includegraphics[width=\textwidth]{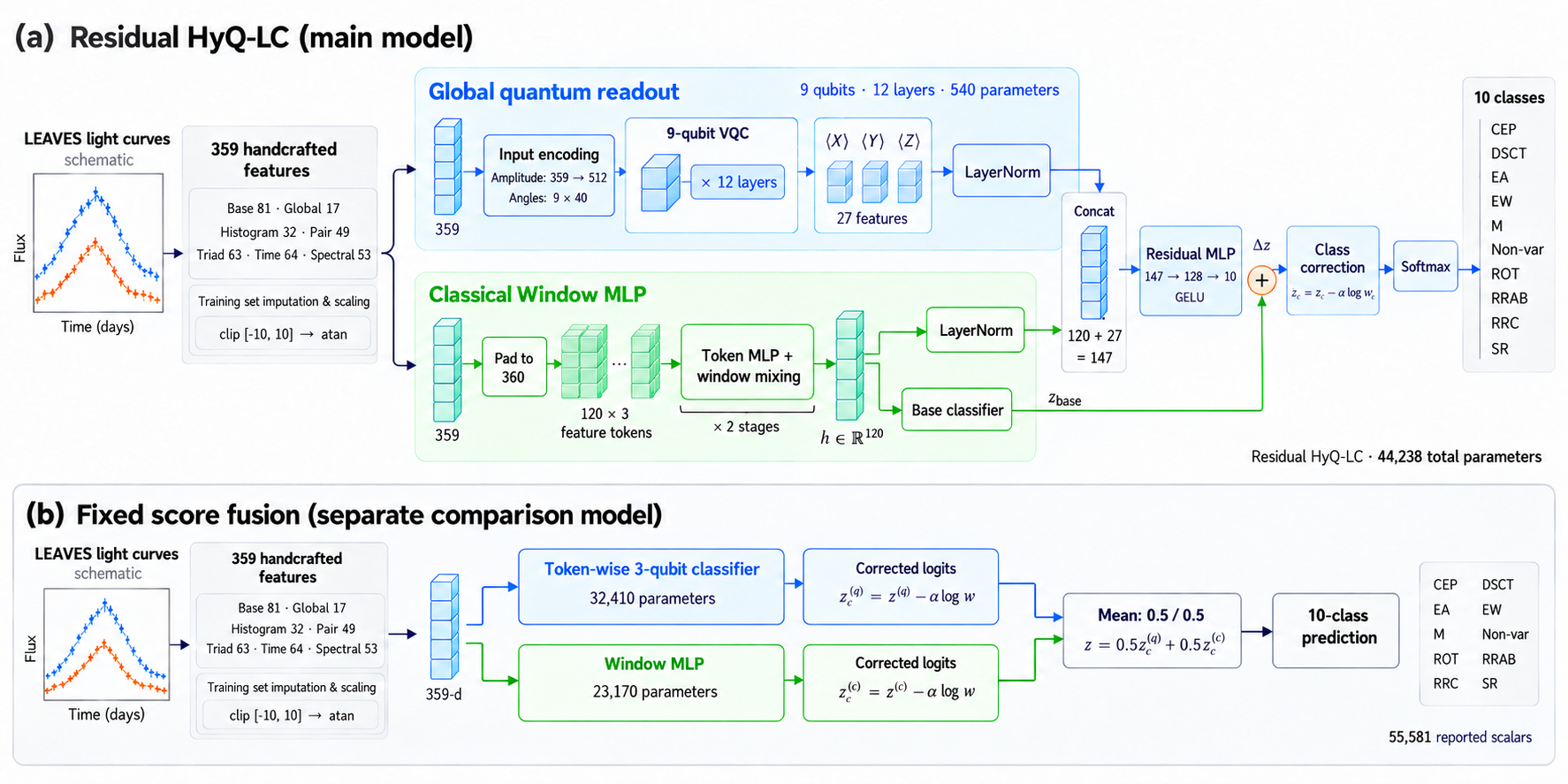}
\caption{HyQ-LC model overview. (a) The main model combines a 120-dimensional Window MLP representation with 27 features from a global nine-qubit readout. Separately normalised features enter a $147\!\to\!128\!\to\!10$ residual MLP, whose logits are added to the backbone prediction before class-weight correction. (b) The separate fixed-fusion configuration averages the corrected logits of independently trained token-wise quantum and Window MLP classifiers. Its 55{,}581-scalar budget includes one fixed mixing coefficient. Light curves are schematic illustrations, not measured examples; the tokens group feature entries rather than time samples.}
\label{fig:arch}
\end{figure}

The main HyQ-LC model combines a classical feature extractor with a global quantum readout (Figure~\ref{fig:arch}a). The backbone is a 23{,}170-parameter window MLP. Its three-feature tokenisation follows Chen and Kuo~\cite{chenkuo2026}; the backbone used here implements token processing and local window mixing with classical MLP blocks. Let $\mathbf{r}=\arctan(\mathbf{x})$ denote the elementwise bounded input. Padding $\mathbf{r}$ with one zero produces 120 three-component tokens. Token MLPs and two stages of six-component window mixing yield a 120-dimensional feature vector and ten base logits:
\begin{equation}
\mathbf{h}=B_{\eta}(\mathbf{r})\in\mathbb{R}^{120},\qquad
\mathbf{z}_{\mathrm{base}}=W_{\mathrm{base}}\mathbf{h}+\mathbf{b}_{\mathrm{base}}.
\label{eq:backbone}
\end{equation}
The quoted backbone parameter count includes its classifier. The quantum branch processes all 359 input features in a single nine-qubit register, and its measured features enter a residual classifier alongside $\mathbf{h}$. Retaining $\mathbf{z}_{\mathrm{base}}$ gives the added branch a direct classical prediction to refine. It also makes the complete-model comparison distinct from a test of the quantum readout alone.

\subsection{Quantum encoding, circuit, and measurement}

\begin{figure}[tbp]
\centering
\includegraphics[width=\textwidth]{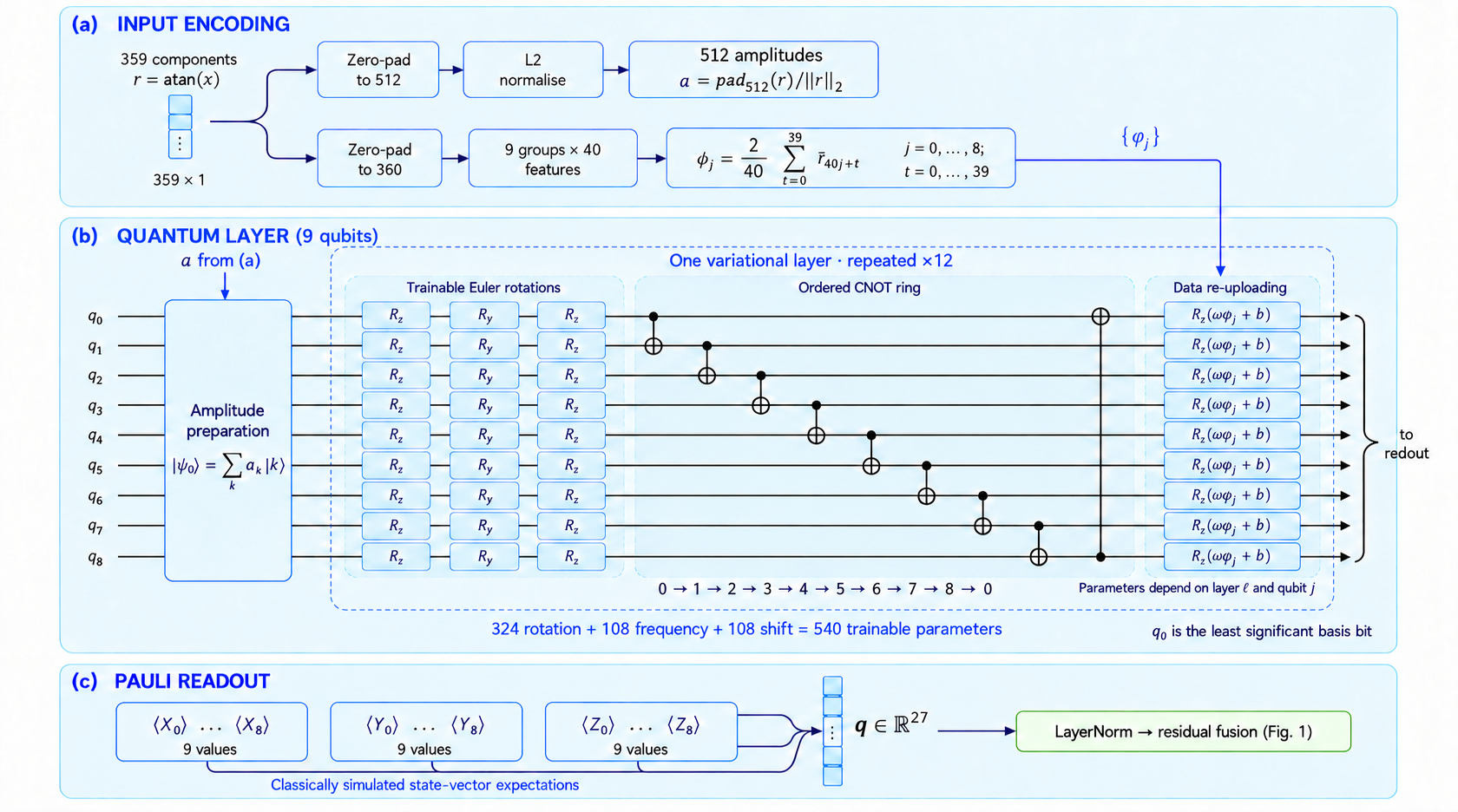}
\caption{Nine-qubit readout in the residual HyQ-LC model. (a) The bounded input supplies 512 normalised amplitudes and nine grouped angles. The amplitude formula shown applies to nonzero inputs; Eq.~\eqref{eq:amplitude} specifies the zero-norm fallback. (b) State preparation is performed once. Each of twelve layers applies trainable $R_z$--$R_y$--$R_z$ rotations, the ordered CNOT ring $0\!\to\!1,\ldots,7\!\to\!8,8\!\to\!0$, and angle re-uploading. The 324 rotation, 108 frequency, and 108 shift parameters sum to 540; qubit zero is the least significant basis bit. (c) Nine expectations for each of $X$, $Y$, and $Z$ form the 27-dimensional output. All expectations are computed by classical state-vector simulation.}
\label{fig:quantum}
\end{figure}

The quantum input combines amplitude encoding with grouped angle re-uploading (Figure~\ref{fig:quantum}a). Define $\bar r_k=r_k$ for $0\leq k<359$ and $\bar r_k=0$ for $359\leq k<512$, and let $\bar{\mathbf r}$ collect these 512 entries. Since $2^9=512$, the encoded state is
\begin{equation}
|\psi_0(\mathbf{x})\rangle=
\begin{cases}
\displaystyle\sum_{k=0}^{511}\frac{\bar r_k}{\|\bar{\mathbf r}\|_2}|k\rangle,
&\|\bar{\mathbf r}\|_2>10^{-8},\\
|0\rangle^{\otimes9},&\text{otherwise}.
\end{cases}
\label{eq:amplitude}
\end{equation}
This encoding uses no learned projection. Normalisation preserves the direction of the transformed feature vector in the amplitudes. The circuit also receives nine angles formed from contiguous groups of 40 padded entries,
\begin{equation}
\varphi_j(\mathbf{x})=\frac{2}{40}\sum_{t=0}^{39}\bar r_{40j+t},
\qquad j=0,\ldots,8.
\label{eq:angles}
\end{equation}
Thus, the grouped angles provide a second input path in addition to the normalised amplitudes. Data re-uploading repeatedly introduces these angles into the trainable circuit~\cite{reuploading}.

The readout has twelve layers (Figure~\ref{fig:quantum}b). In layer $\ell$, each qubit receives an Euler rotation, followed by a ring of CNOT gates and a data-dependent $R_z$ rotation:
\begin{align}
|\psi_\ell\rangle&=D_\ell(\boldsymbol\varphi)\,E\,R_\ell|\psi_{\ell-1}\rangle,
\qquad \ell=1,\ldots,12,\label{eq:circuit}\\
R_\ell&=\prod_{j=0}^{8}R_{z,j}(\theta_{\ell j3})
 R_{y,j}(\theta_{\ell j2})R_{z,j}(\theta_{\ell j1}),\nonumber\\
D_\ell(\boldsymbol\varphi)&=\prod_{j=0}^{8}
 R_{z,j}(\omega_{\ell j}\varphi_j+b_{\ell j}).\nonumber
\end{align}
Here $R_{\nu,j}(t)=\exp(-it\sigma_{\nu,j}/2)$ acts on qubit $j$. The entangler $E$ applies $\mathrm{CNOT}_{j\to(j+1)\bmod9}$ in order $j=0,\ldots,8$, and qubit zero is the least significant basis bit. The trainable $\theta$, $\omega$, and $b$ tensors contain $12\times9\times3=324$, $12\times9=108$, and $108$ parameters, respectively. The complete quantum readout therefore has 540 parameters. Its frequency parameters $\omega$ start at one and its shifts $b$ at zero.

The final state is converted to classical features through single-qubit Pauli expectations (Figure~\ref{fig:quantum}c). Writing $\langle\sigma_j\rangle=\langle\psi_{12}|\sigma_j|\psi_{12}\rangle$, we use
\begin{equation}
\mathbf{q}(\mathbf{x})=
\bigl(\langle X_0\rangle,\ldots,\langle X_8\rangle,
\langle Y_0\rangle,\ldots,\langle Y_8\rangle,
\langle Z_0\rangle,\ldots,\langle Z_8\rangle\bigr)^{\!\top}
\in\mathbb{R}^{27}.
\label{eq:measurement}
\end{equation}
These bounded observables provide a compact representation for the classical classifier~\cite{huang2021}. Input-dependent phases and the entangling gates change the measured features jointly; the classification experiments test whether this particular representation is useful. Encoding affects the accessible function class~\cite{encoding2021}, but circuit depth and expressivity alone do not establish predictive performance~\cite{wang2025}. All expectations are computed by classical state-vector simulation, and gradients are obtained by automatic differentiation through the simulated circuit.

\subsection{Residual fusion and matched classical controls}

The classical and quantum features are separately layer-normalised and concatenated. With $\mathrm{GELU}$ denoting the hidden activation, the final logits are
\begin{align}
\mathbf{u}&=\bigl[\mathrm{LN}_{c}(\mathbf{h});\mathrm{LN}_{q}(\mathbf{q})\bigr]
\in\mathbb{R}^{147},\nonumber\\
\mathbf{z}_{\mathrm{hyb}}&=\mathbf{z}_{\mathrm{base}}+
W_2\,\mathrm{GELU}(W_1\mathbf{u}+\mathbf{b}_1)+\mathbf{b}_2.
\label{eq:residual}
\end{align}
Here $W_1\in\mathbb{R}^{128\times147}$ and $W_2\in\mathbb{R}^{10\times128}$. The final residual layer is initialised to zero, so the added branch initially leaves the backbone logits unchanged. The two normalisation layers and residual MLP contain 20{,}528 parameters. Together with the backbone and quantum readout, they give $23{,}170+540+20{,}528=44{,}238$ parameters.

The pure-MLP control uses the same bounded input, hidden widths 73 and 207, LayerNorm after each hidden linear layer, GELU activations, and a ten-class output. Including biases and normalisation parameters, its total is also 44{,}238. This comparison asks whether the full hybrid configuration improves performance at a fixed parameter count.

A second control targets the readout itself while preserving the residual architecture in Eq.~\eqref{eq:residual}. Its classical input concatenates the first 359 normalised amplitudes from Eq.~\eqref{eq:amplitude} and the nine grouped angles. A fixed sign projection $P\in\{-1,+1\}^{368\times26}$ maps these values to 26 components, followed by
\begin{equation}
\mathbf{q}_{\mathrm{MLP}}=A_2\,\mathrm{GELU}(A_1\mathbf{t}+\mathbf{a}_1),
\quad
\mathbf{t}=P^{\top}\bigl[\mathbf{a}_{0:359};\boldsymbol\varphi\bigr].
\label{eq:matchedreadout}
\end{equation}
The vector $\mathbf{a}_{0:359}$ denotes the first 359 state amplitudes, including the fallback in Eq.~\eqref{eq:amplitude}. With $A_1\in\mathbb{R}^{10\times26}$, $\mathbf{a}_1\in\mathbb{R}^{10}$, and a bias-free $A_2\in\mathbb{R}^{27\times10}$, the readout has $260+10+270=540$ learned parameters. The projection is fixed and contributes no learned parameters. The backbone, normalisation layers, fusion-head initialisation, full parameter count, optimiser-group allocation, and sample order match the quantum configuration. This readout replacement is a more specific test than the complete-model comparison with a pure MLP.

\subsection{Fixed score fusion and tree controls}

A separate fixed-fusion configuration combines a 32{,}410-parameter token-wise quantum classifier with the 23{,}170-parameter window MLP (Figure~\ref{fig:arch}b). The quantum classifier adapts the local three-feature VQC construction of Chen and Kuo~\cite{chenkuo2026} to this task. It is a different configuration from the global nine-qubit residual readout. Let $\tilde{\mathbf z}^{(k)}$ denote the class-corrected logits of classifier $k$, as defined below. The quantum fusion uses
\begin{equation}
\mathbf{z}_{\mathrm{fixQ}}=
\tfrac12\tilde{\mathbf z}^{(\mathrm{tokenQ})}+
\tfrac12\tilde{\mathbf z}^{(\mathrm{window})}.
\label{eq:fixedquantum}
\end{equation}
The classical fusion control combines the window MLP with a 31{,}460-parameter Dense MLP at the probability level:
\begin{equation}
\mathbf{p}_{\mathrm{fixC}}=
0.6\,\mathrm{softmax}(\tilde{\mathbf z}^{(\mathrm{window})})+
0.4\,\mathrm{softmax}(\tilde{\mathbf z}^{(\mathrm{dense})}).
\label{eq:fixedclassical}
\end{equation}
Predictions select the largest fused logit or probability, respectively. Both fusion rules were selected on validation. Reported budgets are 55{,}581 and 54{,}631 scalars, including one fixed mixing coefficient each; each constituent network receives 40 training epochs. Because both the fusion rule and network sizes differ, this is an approximate-budget ensemble comparison.

The LightGBM~\cite{lightgbm} control targets the fixed hybrid's scalar budget using cumulative leaf outputs across all class trees. For $C$ classes, $T$ boosting rounds, and $L$ leaves per tree, the nominal budget is $CTL$. The selected model has $C=10$, $T=370$, and $L=15$, giving 3{,}700 class trees and 55{,}500 leaf values. A separate full-budget reference uses $T=362$ and $L=31$, with 112{,}220 leaf values. Both retain their validation-selected settings. Leaf counts exclude split-node features and thresholds, so they provide a capacity proxy without equating tree representations, neural parameters, or runtime.

\subsection{Training objective and validation selection}

Class imbalance is addressed during neural-network training with inverse-frequency class weights. If $n_c$ is the number of training objects in class $c$ and $N_{\mathrm{tr}}$ is the training-pool size, set $w_c=N_{\mathrm{tr}}/(C n_c)$. For a minibatch $\mathcal B$, the weighted cross-entropy is
\begin{equation}
\mathcal L_{\mathcal B}=-
\frac{\displaystyle\sum_{i\in\mathcal B}w_{y_i}
\log\bigl[\mathrm{softmax}(\mathbf z(\mathbf x_i))\bigr]_{y_i}}
{\displaystyle\sum_{i\in\mathcal B}w_{y_i}}.
\label{eq:loss}
\end{equation}
The weights are computed from training labels only. At validation and inference, logits are adjusted by a separately selected correction strength:
\begin{equation}
\tilde z_c=z_c-\alpha\log w_c,\qquad
\hat y=\mathop{\mathrm{arg\,max}}_{c}\tilde z_c,
\quad\alpha\in\{0,0.25,0.5,0.75,1\}.
\label{eq:correction}
\end{equation}
This correction controls the decision shift associated with class-weighted training. Checkpoints and $\alpha$ are selected by validation Macro-F1. For each pure-MLP seed, the learning rate is selected between $10^{-3}$ and $3\times10^{-3}$ using validation only. ACC in the manuscript tables is evaluated at the same Macro-F1-selected checkpoint.

The pure MLP starts from scratch and receives an 80-epoch budget. The hybrid uses 40 backbone pre-training epochs followed by a 40-epoch branch-training stage. During the first ten branch-training epochs, the backbone learning rate is zero. The selected primary hybrid checkpoints occur at epochs two and four, so their backbone remains at its pre-trained values. The two primary fine-tuning seeds, 42 and 44, vary added weights and sample order while sharing that pre-trained backbone. Thus, the maximum number of training-object presentations is matched, while initialisation, training trajectory, and computational cost differ. The released implementation references specify the optimiser parameter groups and schedules.

\subsection{Evaluation metrics and paired uncertainty}

For each class $c$, let $\mathrm{TP}_c$, $\mathrm{FP}_c$, and $\mathrm{FN}_c$ be true-positive, false-positive, and false-negative counts on the shared test cohort. The two reported metrics are
\begin{equation}
\mathrm{Macro\!\text{-}\!F1}=\frac1C\sum_{c=1}^{C}
\frac{2\mathrm{TP}_c}{2\mathrm{TP}_c+\mathrm{FP}_c+\mathrm{FN}_c},
\qquad
\mathrm{ACC}=\frac{\sum_{c=1}^{C}\mathrm{TP}_c}{N_{\mathrm{test}}},
\label{eq:metrics}
\end{equation}
with $N_{\mathrm{test}}=20{,}000$. Macro-F1 weights every class equally, whereas ACC weights every object equally. They therefore address different aspects of classification on the imbalanced cohort. Table values are percentages, and hybrid-minus-control differences use percentage points.

Uncertainty is evaluated with 10{,}000 ordinary paired-object bootstrap resamples. For metric $M$, the difference in resample $b$ is
\begin{equation}
\Delta_M^{*(b)}=100\!\left[
M(\mathbf y_{I_b},\hat{\mathbf y}^{Q}_{I_b})-
M(\mathbf y_{I_b},\hat{\mathbf y}^{C}_{I_b})\right],
\label{eq:bootstrap}
\end{equation}
where $I_b$ contains $N_{\mathrm{test}}$ indices sampled with replacement. The same indices are used for both predictions, preserving object-level pairing; class proportions are allowed to fluctuate. The 2.5th and 97.5th percentiles give the reported 95\% intervals. These intervals condition on the fitted models and exclude training-seed and model-search variation; no multiplicity correction is applied. The release also includes additional seeded runs and numerical ablations.

\section{Results}

\subsection{Comparison with the parameter-matched pure MLP}
\label{sec:sameparameter}

The residual hybrid had higher Macro-F1 and accuracy than the pure MLP on both primary seeds (Table~\ref{tab:puremlp}). The F1 interval included zero for seed~42 and was positive for seed~44; both ACC intervals were positive. These differences compare the complete trained configurations, including their different architectures and pre-training histories.

\begin{table}[htbp]
\centering
\small
\setlength{\tabcolsep}{4pt}
\caption{Parameter-matched comparison on 20{,}000 test objects. Both models have 44{,}238 parameters. F1 scores are percentages; differences are in pp and use unrounded scores. Brackets give paired-bootstrap 95\% confidence intervals.}
\label{tab:puremlp}
\begin{tabular}{@{}cccccc@{}}
\toprule
Seed & Hybrid F1 & MLP F1 & $\dF$ (pp) & F1 95\% interval & $\dA$ (pp), 95\% interval \\
\midrule
42 & 92.2913 & 91.4637 & $+0.8276$ & $[-0.0650,+1.6945]$ & $+0.6000$ $[+0.3450,+0.8550]$ \\
44 & 92.3005 & 91.0611 & $+1.2393$ & $[+0.5153,+2.0053]$ & $+0.8450$ $[+0.5900,+1.1000]$ \\
\bottomrule
\end{tabular}
\end{table}

\subsection{Replacing the quantum branch}

Within the shared residual architecture, the quantum readout had slightly higher F1 point estimates than the matched classical readout (Table~\ref{tab:replacements}). Both intervals included zero. This test therefore leaves the predictive contribution specific to the quantum branch unresolved.

\begin{table}[htbp]
\centering
\small
\setlength{\tabcolsep}{4pt}
\caption{Readout replacement within the residual hybrid on the same test objects. Both readouts have 540 parameters and both full models have 44{,}238 parameters. Scores are Macro-F1 (\%); differences and intervals are in pp.}
\label{tab:replacements}
\begin{tabular}{@{}ccccc@{}}
\toprule
Seed & Quantum readout & MLP readout & $\dF$ (pp) & 95\% interval \\
\midrule
42 & 92.2913 & 92.2708 & $+0.0205$ & $[-0.3299,+0.3726]$ \\
44 & 92.3005 & 92.0039 & $+0.2965$ & $[-0.0333,+0.6982]$ \\
\bottomrule
\end{tabular}
\end{table}

\subsection{Fixed fusion and tree controls}

The fixed quantum fusion achieved 92.3371\% Macro-F1 and 94.7450\% ACC (Table~\ref{tab:leafbudget}). It had a small F1 point gain over classical fusion, with essentially unchanged ACC. Against leaf-budget-matched LightGBM, its F1 was $0.1166\pp$ higher but ACC was $0.1200\pp$ lower, corresponding to 24 fewer correct classifications. The paired F1 intervals spanned zero for both controls. The full-budget LightGBM reference scored higher on both metrics.

The classwise comparison (Table~\ref{tab:classwise}) shows the largest positive F1 differences against the budget-matched tree in CEP ($+1.0040\pp$), RRC ($+1.1429\pp$), and ROT ($+0.5128\pp$). CEP and RRC contain only 45 and 357 test objects. Equal class weighting allows these gains to raise Macro-F1 even when total correct predictions decrease. This is a class-specific pattern, not an improvement across all minority classes.

\begin{table}[htbp]
\centering
\small
\setlength{\tabcolsep}{4pt}
\caption{Controls for fixed quantum fusion on the same 20{,}000 test objects. The hybrid reference scores F1/ACC of 92.3371/94.7450\%. Control scores are percentages; hybrid-minus-control differences and paired 95\% intervals are in pp.}
\label{tab:leafbudget}
\begin{tabular}{@{}lccc@{}}
\toprule
Control & F1/ACC (\%) & $\dF$ (pp), 95\% interval & $\dA$ (pp) \\
\midrule
Classical fusion & 92.1510 / 94.7500 & $+0.1860$ $[-0.2869,+0.6740]$ & $-0.0050$ \\
Budget-matched LightGBM & 92.2204 / 94.8650 & $+0.1166$ $[-0.8111,+1.0763]$ & $-0.1200$ \\
Full-budget LightGBM & 92.9681 / 95.1100 & $-0.6311$ $[-1.4192,+0.1197]$ & $-0.3650$ \\
\bottomrule
\end{tabular}
\end{table}

\begin{table}[htbp]
\centering
\small
\setlength{\tabcolsep}{5pt}
\caption{Class distribution and classwise F1 on the same 20{,}000 test objects. Quantum fusion denotes the fixed-fusion configuration; LightGBM is the 55{,}500-leaf-value control. Shares and F1 scores are percentages; $\dF$ is quantum fusion minus LightGBM in pp. The last row gives total counts and unweighted mean F1.}
\label{tab:classwise}
\begin{tabular}{@{}lrrrrr@{}}
\toprule
Class & Objects & Share (\%) & Quantum F1 & LightGBM F1 & $\dF$ (pp) \\
\midrule
CEP & 45 & 0.225 & 84.3373 & 83.3333 & $+1.0040$ \\
DSCT & 337 & 1.685 & 86.3636 & 86.4536 & $-0.0899$ \\
EA & 912 & 4.560 & 96.8097 & 97.1239 & $-0.3142$ \\
EW & 6{,}307 & 31.535 & 94.7460 & 94.7427 & $+0.0034$ \\
M & 352 & 1.760 & 97.5955 & 97.5887 & $+0.0068$ \\
Non-var & 2{,}566 & 12.830 & 96.0356 & 96.7356 & $-0.7000$ \\
ROT & 2{,}257 & 11.285 & 86.3818 & 85.8691 & $+0.5128$ \\
RRAB & 796 & 3.980 & 97.5579 & 97.7358 & $-0.1779$ \\
RRC & 357 & 1.785 & 86.0908 & 84.9478 & $+1.1429$ \\
SR & 6{,}071 & 30.355 & 97.4523 & 97.6740 & $-0.2217$ \\
\midrule
Total / Macro-F1 & 20{,}000 & 100.000 & 92.3371 & 92.2204 & $+0.1166$ \\
\bottomrule
\end{tabular}
\end{table}

\FloatBarrier
\section{Discussion and Conclusion}

The comparisons separate performance of the complete classifier from evidence about its quantum component. The residual hybrid has higher test-score point estimates than the tested parameter-matched pure MLP, yet its shared-architecture readout replacements leave the quantum-specific increment unresolved. Together, these results support the complete configuration as a useful candidate while leaving architecture, pre-training, and branch representation as competing explanations for its gains. This distinction follows the benchmark concerns raised by Bowles et al.~\cite{bowles2024} and complements architecture-level comparisons such as Chen and Kuo~\cite{chenkuo2026}.

The classwise results explain why model ranking depends on the evaluation criterion. Under Eq.~\eqref{eq:metrics}, the F1 score of CEP receives the same weight as that of EW despite their very different sample sizes. The fixed hybrid's improvements in CEP, RRC, and ROT can therefore offset declines in other classes at the Macro-F1 level while total correct predictions decrease. This makes the result relevant to class-balanced classification, but it does not establish a uniform improvement for rare objects. In particular, the CEP estimate rests on only 45 test objects, and classwise F1 changes combine precision and recall rather than measuring recovered objects alone.

The tree comparisons further limit what can be inferred from a scalar budget. Similar leaf-output and neural-parameter counts do not imply equivalent function classes, optimisation effort, or inference cost. The full-budget LightGBM reference scores higher on both metrics, showing that the small F1 advantage over its budget-matched counterpart does not persist across tree budgets. Likewise, the different logit and probability mixtures in Eqs.~\eqref{eq:fixedquantum} and~\eqref{eq:fixedclassical} prevent attributing their ensemble difference solely to the quantum constituent.

The remaining questions concern generalisation and the source of the residual gain. Evaluation on previously unexamined objects would test persistence beyond the explored cohort. Independently pre-trained backbones would distinguish fine-tuning variation from full training variation, while matched fusion rules would sharpen the ensemble comparison. These tests would extend the current evidence rather than change the reported conditional comparisons. All circuits here are classically simulated; neither a quantum-hardware speed advantage nor a general quantum advantage follows from the present results.

\section{Data and Code Availability}

The repository provides the shared test features, labels, per-object predictions, result tables, and a portable metric-recomputation script~\cite{artifacts}. Checkpoint and source-prediction hashes document provenance. Standalone quantum-readout code and original training implementation references are included; the latter require the authors' cluster environment. The release supports independent recomputation of the reported statistics from saved predictions.

\begin{acknowledgments}
The authors thank the developers of PyTorch, LightGBM, and PennyLane~\cite{pennylane}.
\end{acknowledgments}


\begin{thebibliography}{99}
\footnotesize
\bibitem{moriwaki2023} K.~Moriwaki, T.~Nishimichi, and N.~Yoshida, \emph{Machine learning for observational cosmology}, \href{https://doi.org/10.1088/1361-6633/acd2ea}{Reports on Progress in Physics \textbf{86}, 076901 (2023)}.
\bibitem{alerce2021} P.~S\'anchez-S\'aez \emph{et al.}, \emph{Alert Classification for the ALeRCE Broker System: The Light Curve Classifier}, \href{https://doi.org/10.3847/1538-3881/abd5c1}{The Astronomical Journal \textbf{161}, 141 (2021)}.
\bibitem{tdlight2026} X.~Yu, C.~Yu, Z.~Shao, C.~Wang, and B.~Yang, \emph{TDLight: A Framework for Incremental Light-curve Management and Continuous Classification Optimization}, \href{https://doi.org/10.3847/1538-4365/ae8300}{The Astrophysical Journal Supplement Series \textbf{285}, 57 (2026)}.
\bibitem{astromer2023} C.~Donoso-Oliva \emph{et al.}, \emph{ASTROMER: A transformer-based embedding for the representation of light curves}, \href{https://doi.org/10.1051/0004-6361/202243928}{Astronomy \& Astrophysics \textbf{670}, A54 (2023)}.
\bibitem{starembed2026} W.~Li \emph{et al.}, \emph{StarEmbed: Benchmarking Time Series Foundation Models on Astronomical Observations of Variable Stars}, \href{https://proceedings.mlr.press/v306/li26t.html}{Proceedings of the 43rd International Conference on Machine Learning, PMLR \textbf{306}, 66289--66329 (2026)}.
\bibitem{shao2026} Z.~Shao, C.~Yu, B.~Yang, and X.~Chen, \emph{Reframing Variable Star Classification as Structured Sequence Modeling with Large Language Models}, \href{https://doi.org/10.3847/1538-4357/ae8ade}{The Astrophysical Journal \textbf{1007}, 82 (2026)}.
\bibitem{mari2020} A.~Mari, T.~R.~Bromley, J.~Izaac, M.~Schuld, and N.~Killoran, \emph{Transfer learning in hybrid classical-quantum neural networks}, \href{https://doi.org/10.22331/q-2020-10-09-340}{Quantum \textbf{4}, 340 (2020)}.
\bibitem{peters2021} E.~Peters \emph{et al.}, \emph{Machine learning of high dimensional data on a noisy quantum processor}, \href{https://doi.org/10.1038/s41534-021-00498-9}{npj Quantum Information \textbf{7}, 161 (2021)}.
\bibitem{chenkuo2026} C.-S.~Chen and E.-J.~Kuo, \emph{Do Quantum Transformers Help? A Systematic VQC Architecture Comparison on Tabular Benchmarks}, \href{https://arxiv.org/abs/2604.23931v2}{arXiv:2604.23931v2 (2026)}.
\bibitem{bowles2024} J.~Bowles, S.~Ahmed, and M.~Schuld, \emph{Better than classical? The subtle art of benchmarking quantum machine learning models}, \href{https://arxiv.org/abs/2403.07059}{arXiv:2403.07059 (2024)}.
\bibitem{leaves2024} Y.~Fei \emph{et al.}, \emph{LEAVES: An Expandable Light-curve Data Set for Automatic Classification of Variable Stars}, \href{https://doi.org/10.3847/1538-4365/ad785b}{The Astrophysical Journal Supplement Series \textbf{275}, 10 (2024)}.
\bibitem{leavesdata} C.~Yu, \emph{An Expandable Light Curve Dataset for Automatic Classification of Variable Stars (LEAVES)}, National Astronomical Data Center of China (2024), \url{https://doi.org/10.12149/100962}.
\bibitem{artifacts} Project artifacts: data, code, audited results, and verification scripts, \url{https://github.com/bestdo77/quantum-lightcurve-hybrid} (2026).
\bibitem{reuploading} A.~P\'erez-Salinas, A.~Cervera-Lierta, E.~Gil-Fuster, and J.~I.~Latorre, \emph{Data re-uploading for a universal quantum classifier}, \href{https://doi.org/10.22331/q-2020-02-06-226}{Quantum \textbf{4}, 226 (2020)}.
\bibitem{huang2021} H.-Y.~Huang \emph{et al.}, \emph{Power of data in quantum machine learning}, \href{https://doi.org/10.1038/s41467-021-22539-9}{Nature Communications \textbf{12}, 2631 (2021)}.
\bibitem{encoding2021} M.~Schuld, R.~Sweke, and J.~J.~Meyer, \emph{Effect of data encoding on the expressive power of variational quantum-machine-learning models}, \href{https://doi.org/10.1103/PhysRevA.103.032430}{Physical Review A \textbf{103}, 032430 (2021)}.
\bibitem{wang2025} X.~Wang, H.~Tao, and R.~Wu, \emph{Predictive Performance of Deep Quantum Data Re-uploading Models}, \href{https://proceedings.mlr.press/v267/wang25cw.html}{Proceedings of the 42nd International Conference on Machine Learning, PMLR \textbf{267}, 64491--64524 (2025)}.
\bibitem{lightgbm} G.~Ke \emph{et al.}, \emph{LightGBM: A Highly Efficient Gradient Boosting Decision Tree}, \href{https://proceedings.neurips.cc/paper_files/paper/2017/hash/6449f44a102fde848669bdd9eb6b76fa-Abstract.html}{Advances in Neural Information Processing Systems \textbf{30}, 3146--3154 (2017)}.
\bibitem{pennylane} V.~Bergholm \emph{et al.}, \emph{PennyLane: Automatic differentiation of hybrid quantum-classical computations}, \href{https://arxiv.org/abs/1811.04968}{arXiv:1811.04968 (2018)}.
\end{thebibliography}
\end{document}